\documentstyle[11pt,newpasp,twoside,epsf]{article}
\def\edcomment#1{\iffalse\marginpar{\raggedright\sl#1\/}\else\relax\fi}
\reversemarginpar

\begin{document}
\title{The Nature of the Local Group}
 \author{Antonio Aparicio}
\affil{University of La Laguna and Instituto de Astrof\'\i sica de Canarias,
  V\'\i a L\'actea, E-38205 La Laguna, Tenerife, Spain}

\begin{abstract}
The Local Group provides an interesting and representative sample of
  galaxies in the rest of the Universe. The high accuracy with which many
  problems can be addressed in Local Group galaxies is of paramount importance
  for understanding galaxy formation and evolution. This contribution
  presents a short review of overall Local Group properties followed by short
  discussions of five topics in which the study of Local Group members
  provides particularly significant information. These topics are only
  examples of the usefulness and potential of Local Group research. The five
  selected topics are the formation of the Milky Way, galaxy destruction and tidal
  streams, detailed galactic chemical evolution, star formation history
  determination, and low surface brightness extended structures.

\end{abstract}

\section{Introduction}

The main motivation for studying the Local Group is that it is our nearest sample
of galaxies. This would be a rather weak reason if we
could not assume that the Local Group galaxies are a good, unbiased,
representative sample of the galaxies in the rest of the universe. Only in
such a case could the more accurate results that we can obtain for the Local Group
members be extended to other galaxies. In fact, the Local Group
contains members of most galaxy types: big design spirals (Andromeda, the Milky
Way, M 33); luminous irregulars (the Large and Small Magellanic Clouds); dim
irregulars (NGC 6822, IC 1613, Leo A, etc.); a possible blue compact dwarf (IC
10); a nucleated dwarf elliptical (M 32); dwarf ellipticals (NGC 205, NGC
147, NGC 185); dwarf spheroidals (Sculptor, Ursa Minor, etc.), and a highly
stripped galaxy (Sagittarius dSph). The only important types of galaxies not
present in the Local Group are ellipticals (non-dwarfs) and cDs (see van den
Bergh 2000).

In this contribution, after an overview of Local Group properties, I will
review a few general topics with a common characteristic: that the study of
Local Group members should provide a significant improvement in our knowledge
of the problem. The topics I have selected are the following:

\vspace {2mm}
$\bullet$ Galaxy formation: the Milky Way

\vspace {2mm}
$\bullet$ Galaxy destruction and tidal streams: the Sagittarius dSph galaxy

\vspace {2mm}
$\bullet$ Age-metallicity laws: detailed chemical evolution

\vspace {2mm}
$\bullet$ Quantitative determinations of star formation histories

\vspace {2mm} $\bullet$ Extended low surface brightness structures in dwarf
galaxies

\vspace {2mm}
The properties of the Local Group and its galaxies have recently been 
reviewed in several meetings ({\it The Local Group: Comparative and Global
  Properties} (Layden, Smith, \& Storm 1994); {\it The Stellar Content of the
  Local Group Galaxies} (Whitelock \& Cannon 1999) and in the Canary Islands
Winter School {\it Stellar Astrophysics for the Local Group} (Aparicio,
Herrero \& S\'anchez 1998). Recent fundamental monographic reviews have been given by
Grebel (1997), Mateo (1998), and van den Bergh (2000).

\section{The Local Group: An Overview}

The number of Local Group galaxy members has been increasing since Hubble (1936) introduced the term "Local Group" itself. Hubble
mentioned only ten galaxies as members of the Local Group: the Milky Way, the Large
Magellanic Cloud (LMC), the Small Magellanic Cloud (SMC), Andromeda, M 32, NGC
205, M 33, NGC 6822, IC 1613, and IC 10 (which Hubble did not consider a
secure member). Since then, galaxies have been discovered continuously
and the rate of discover shows no sign of abating for the moment (see
the enlightening figure 1.1 in van den Bergh 2000). 

Deciding whether or not a galaxy is a Local Group member is not a straightforward task.  Certainly the best
criterion is dynamical: member galaxies should be gravitationally bound
to the Local Group. But i) the mass of the Local Group is not reliably known and ii) the
tangential velocity of galaxies cannot yet be measured (except for the very
nearest ones). Using an estimate of the Local Group mass, Sandage (1986)
found a dynamical solution implying that galaxies having velocities in excess
of 60 km s$^{-1}$ relative to the Local Group barycenter should be unbound.  The
problem has been several times revisited.  In particular, Karachentsev \&
Makarov (2001) determined the radius, $R_0$ of the zero velocity sphere,
which separates the Local Group from the overall cosmological expansion,
obtaining $R_0=0.96\pm 0.05$ Mpc.  Lacking more accurate information, it is a
reasonable and common choice to consider galaxies within this radius as
probable members of the Local Group.

Summarizing, the current list of Local Group members or possible members
includes about 35 to 40 galaxies. Eight of these have been discovered or
identified as Local Group members or possible members since 1990: Sextans
(Irwin et al.\ 1990); Tucana (Corwin et al.\ 1985, discovery; Lavery 1990,
identification as member); Antlia (Corwin et al.\ 1985, discovery;
Whiting et al.\ 1997, identification as member); Sagittarius
(Ibata et al.\ 1994); Andromeda V (Armandroff et al.\ 1998); Andromeda VI (also
called Pegasus II, Grebel, \& Guhathakurta 1999), Andromeda VII (also called
Cassiopeia, Grebel, \& Guhathakurta 1999) and Andromeda VIII (Morrison et al.\
2003). The continuous discovery of Local Group galaxies includes Milky Way
satellites, two of which (Sextans and Sagittarius) were discovered in
the nineties. Figure 1 shows the rate of these discoveries.  Nothing
indicates that the discovery process is now finished.

Figure 2 shows a general perspective of the Local Group. Figure 3 is a zoom
of the former showing the Milky Way and Andromeda subsystems in more detail.
Figures 4 and 5 shows the Andromeda and Milky way subsystems as seen from
the Milky Way and Andromeda, respectively. The concentration of dwarf
spheroidal galaxies around the two big spirals is clear in all the cases as
well as the distribution of irregulars filling the Local Group field.

\section{Galaxy Formation: The Milky Way}

The first case that we will snapshot in which the study of Local Group members is
particularly enlightening is the Milky Way formation process.  The formation
timescale of the Milky Way has been the subject of debate for several years.
The two milestone papers representing this debate are those by Eggen,
Lynden-Bell, \& Sandage (1962) and Searle \& Zinn (1978). In the last decade,
Chaboyer, Demarque, \& Sarajedini (1996), from an analysis of the ages of a
globular cluster sample, proposed that the formation of the Milky Way halo
lasted 5 Gyr, or 30--40\% of the age of the Galaxy. A more accurate analysis based on
some 50 globular cluster, in which major attention was paid to the
homogeneity of the data sample, is being carried on by Rosenberg and
collaborators. Rosenberg et al.\ (1999) published a first set of results
corresponding to the inner 20 kpc. A further extension of the work (Rosenberg
et al.\ 2004) to the external halo (40 kpc) indicates that perhaps two
separate globular cluster populations are present, one of them 20\%
younger than the other (Figure 6). Clusters are coeval (within the errors) in
each group, the oldest one containing about 70\% of the objects. Besides these
two components a few still younger clusters exist. These are
represented by only two objects in the Rosenberg et al.\ (2004) sample. 

\section{Galaxy Destruction and Tidal Streams: The Sagittarius dSph}

The discovery of the Sagittarius dSph (Ibata et al.\ 1994) and of the fact
that it is in an advanced state of tidal disruption have been of paramount
relevance in the galaxy interaction research field. In particular, the
Sagittarius dwarf is a living example of a fundamental process of galaxy
evolution: the disruption and merging of dwarf galaxies in the potential well
of giants, which is one of the predictions of the hierarchical galaxy
formation scenario.  The Sagittarius dwarf extends over a stream that
completely enwraps the Milky Way in a polar orbit. The accuracy reached by
several searches and positive and negative detections of the Sagittarius
stream at different galactic latitudes is making it possible to develop an
accurate orbital model for it and to constrain the Milky Way potential well.
Figure 7 shows the orbital model for the Sagittarius stream by Mart\'\i
nez-Delgado, G\'omez-Flechoso, \& Aparicio (2003). The center of the galaxy
marked in the figure corresponds to the globular cluster M 54. Other globular
clusters are also marked. One of them, Pal 12, is particularly young.

\section{Age-Metallicity Laws: Detailed Chemical Evolution}

A knowledge of the chemical enrichment process is fundamental to a proper
understanding of the evolution of galaxies. In the simple closed-box model, it only
depends on the star formation rate (SFR) as a function of time. Determining
the SFR is a complicated task, but things made are still more difficult because
gas infall into and outflow from the intergalactic medium also play a
quite fundamental role. Chemical enrichment laws are often assumed in
astrophysical research, but they have only very recently obtained observationally. An example is the Fornax dSph galaxy (Pont et al.\ 2003).
This galaxy is situated 138 kpc from the Milky Way. At this distance it is
Possible i) to measure the metallicity of upper red giant branch
(RGB) stars from the Ca II triplet and ii) to determine age of the
same stars from a direct comparison with the isochrones of the corresponding
metallicity. The latter (and perhaps the former) produces inaccurate absolute
scales. But the internal accuracy is high. Figure 8 shows a metallicity--age
plot for the star sample observed in Fornax. The dot distribution can be
matched to metallicity laws derived from several enrichment scenarios. The
one shown corresponds to the best solution for infall and outflow rates,
assuming a simple case of constant SFR (Pont et al.\ 2003).

\section{Quantitative Determinations of Star Formation Histories}

Galaxies evolve along two main paths: dynamically, including interaction with
external systems, and through the process of the formation, evolution, and
death of the stars within them. The latter has the following relevant effects
on the galaxy: i) the evolution of gas content, ii) chemical
enrichment, and iii) the formation of the stellar populations. The star
formation history (SFH), is therefore fundamental to understanding the galaxy
evolution process. We define the SFH as composed of two simpler functions:
the SFR as a function of time and the chemical evolution law. The
explicit determination of the SFH requires direct information about age and
metallicity of the stars belonging to successive stellar populations. The
color--magnitude diagram (CMD) provides this kind of information for large
star samples and is the best tool for deriving the SFH. But for this
information to be useful, the CMD must be deep and accurate, which is only
reachable in the Local Group or its neighborhood.

Deriving a quantitative, accurate SFH is, in any case, complicated and
require a more or less sophisticated technique. The standard procedure
involves three main ingredients: i) good data, from which a deep
observational CMD, ideally reaching the oldest main-sequence turn-offs, can
be plotted; ii) a stellar evolution library providing colors and magnitudes
of stars as a function of age, mass, and metallicity, and iii) a method for
relating the number of stars populating different regions of the
observational CMD with the density distribution of stars in the CMD as a
function of age, mass, and metallicity, as predicted by the stellar evolution
models. The latter will provide the SFH of the system. A review of the procedures is given by Aparicio (2003).

The CMD depth is fundamental if an SFH extending to the early epoch of 
galaxy evolution as a star-forming system is required. For CMDs showing only
the RGB and the upper main sequence and blue loops, only an integrated
estimate of the SFR is possible for ages older than a few hundred million
years. In figure 6 of the review by Aparicio (2003), the SFRs of
the galaxies are plotted for which, at the time of writing the paper in 2001, a
quantitative study of the CMD had been done using synthetic CMD techniques.
In these cases the SFR was derived for the entire for the entire 
galaxy lifetime. Other,
less accurate or less time extended SFRs are shown in figure 7 of the same
paper.

With accurate enough data, the simultaneous derivation of the SFR and the CEL
is possible, providing the full SFH as we have previously defined it. Figure
9 shows an example of that for the dwarf galaxy Phoenix (Hidalgo,
Aparicio, \& Mart\'\i nez-Delgado 2004). A plot of this kind, in which the
SFR and the CEL are plotted as a function of time, is customarily called a
``population box'' (Hodge 1989).

\section{Extended Low Surface Brightness Structures in Dwarf
Galaxies}

Dekel \& Silk (1986) proposed that the driving winds of supernovae (SNe) of
the first stellar generation should have swept out most of the gas in dwarf
galaxies. Those retaining some of it would eventually be able to produce new
stellar generations, resulting in dIrrs. In such cases, dIrrs should still
bear the traces of their first stellar generation, which would be seen today
as an extended structure populated by very old low metallicity stars.
Extended structures lacking young stars are being routinely discovered in
dIrrs (e.g.,  Minniti, \& Zjilstra 1996; Aparicio et al.\ 2000). However, it
is not yet clear whether the stellar content of these structures is a really
old, low metallicity primeval population or the superposition of successive
old and intermediate-age populations that simply lack young stars. Recent
results even indicate that these extended structures could contain a
significant young population. Particularly interesting is the work by
Komiyama et al.\  (2003) on the dwarf galaxy NGC 6822. This galaxy was known
to have a gas component much more extended than the conspicuous optical
component (Roberts 1972; de Blok \& Walker 2000). These very extended gas
structures are common in gas rich dwarf galaxies (see Roberts 1972 for several
examples). Komiyama et al.\ (2003), from wide-field observations obtained
with the Subaru Telescope, have obtained a map of young blue stars, as
shown in their figure 4. The distribution of young stars
matches the extended gas component reasonably well. Even more, the young
stars seem to trace a very low surface brightness spiral arm structure. The
high surface brightness body of NGC 6822, which had been considered the
only stellar component of the galaxy turns out to be only its central component.

On another hand, dynamical models published by Mayer et al.\ (2001; see their
figure 12) show that, evolving within a big galaxy potential well, high
surface brightness galaxies (as dwarf irregulars) develop a low surface brightness
spiral structure before becoming dwarf elliptical galaxies. This could
indicate that galaxies like NGC 6822 could be in a transitional phase toward a
dwarf elliptical morphology.

\section{Final Considerations}

In this paper I have reviewed only a few relevant topics related to Local Group
astrophysics. There are, however, several fields of astrophysical research, ranging from
stellar astrophysics to cosmology, that could benefit from study of Local Group. In
his book {\it The Realm of Nebulae}, Hubble (1936) pointed out that the study
of individual galaxies in the Local Group was important because 1) they were
`the nearest and most accessible examples of their particular types' and
2) they provided `a sample collection of nebulae, from which criteria can
be derived for further exploration' (I take these quotations from van den
Bergh 2000). Hubble's motivation continues to be valid today.

\acknowledgments

I am grateful to Drs Gallart, Pont, Hidalgo, Rosenberg, Mart\'\i
nez-Delgado, and G\'omez-Flechoso for fruitful discussions and comments that
have largely improved this paper, and to Drs Pont, Zinn, Gallart, Hardy, and
Winnick for having made the data of one of their plots available to me
(Figure 8 of this paper).

\newpage

\begin{figure}[!h]
\caption{The Milky Way stellite system discovery. The number of known Milky
  Way satellites is plotted against the year. The plot seems to indicate that
  the discovering era is not yet finished.}
\end{figure}

\begin{figure}[!h]
\caption{The Local Group in perspective. Labels on the concentric circles are 
  distances to the Local Group barycenter in Mpc. The two galaxy subsystems
  around the Milky Way and Andromeda are visible}
\end{figure}

\begin{figure}[!h]
\caption{The central region of the Local Group in perspective showing in
  detail the Milky Way and Andromeda subsystems. Labels on the concentric
  circles are distances to the Local Group barycenter in Mpc.}
\end{figure}

\begin{figure}[!h]
\caption{The Andromeda subsystem as seen from the Milky Way}
\end{figure}

\begin{figure}[!h]
\caption{The Milky Way subsystem as seen from the Andromeda}
\end{figure}

\begin{figure}[!h]
\caption{Age of Milky Way halo globular clusters as a function of
  galactocentric distance. Open dots correspond to an extended sample and
  are preliminary (Rosenberg et al.\ 2004).}
\end{figure}

\begin{figure}[!h]
\caption{Dynamical model from the Sagittarius stream from Mart\'\i
  nez-Delgado et al.\ (2003). The center of the galaxy, marked by a square,
  corresponds to the position of the M 54 globular cluster. The positions of
  the globular clusters Pal 2, Pal 12, and Pal 14 are also marked. Other
  symbols correspond to several detections of the galaxy. See
  Mart\'\i nez-Delgado et al.\ (2003) for references.}
\end{figure}

\begin{figure}[!h]
\caption{The metallicity law of the Fornax dwarf galaxy. Dots correspond to the age--metallicity pairs measured for the
  sample stars. The fit reproduces the metallicity law
  corresponding to the best solution for star formation, infall, and outflow
  scenario (Pont et al.\ 2003).}
\end{figure}

\begin{figure}[!t]
\caption{The population box of the dwarf galaxy Phoenix. Solutions of this 
  kind, in which both the SFR and the CEL are simultaneously derived as a
  function of time, are possible for deep and accurate CMDs (Hidalgo \&
  Aparicio 2004). This result is preliminary.}
\end{figure}

\end{document}